\documentclass[%
 reprint,
superscriptaddress,
showpacs,preprintnumbers,
 amsmath,amssymb,
 prb,
]{revtex4-1}

\usepackage{graphicx}
\usepackage{dcolumn}
\usepackage{bm}
\usepackage{xcolor}

\begin{document}

\newcommand{\Cs}{CsFe$_2$As$_2$}
\newcommand{\K}{KFe$_2$As$_2$}
\newcommand{\Tc}{$T_{\rm c}$}
\newcommand{\Pc}{$P_{\rm c}$}
\newcommand{\ie}{{\it i.e.}}
\newcommand{\eg}{{\it e.g.}}
\newcommand{\etal}{{\it et al.}}  
\newcommand{\mucm}{$\mathrm{\mu\Omega\, cm}$}

\newcommand{\YRS}{YbRh$_2$Si$_2$}
\newcommand{\Co}{CeCoIn$_5$}
\newcommand{\Rh}{CeRhIn$_5$}
\newcommand{\p}[1]{\left(#1 \right)}
\newcommand{\Tn}{$T_{\rm N}$}

\newcommand{\Tmin}{$T_{\rm min}$}
\newcommand{\Tstar}{$T^\star$}
\newcommand{\Hvs}{$H_{\rm vs}$}
\newcommand{\Hc}{$H_{\rm c2}$}
\newcommand{\Hcstar}{$H_{\rm c2}^\star$}
\newcommand{\Hstar}{$H^\star$}
\newcommand{\Nqp}{$N_{\rm qp}$}
\newcommand{\Nsc}{$N_{\rm sc}$}
\newcommand{\NbSi}{Nb$_{0.15}$Si$_{0.85}$}
\newcommand{\RH}{$R_{\rm H}$}
\newcommand{\xc}{$x_{\rm c}$}
\newcommand{\LB}{$\ell_{\rm B}$}


\title{Sudden reversal in the pressure dependence of \Tc~in the iron-based superconductor CsFe$_2$As$_2$:
A possible link between inelastic scattering and pairing symmetry}

\author{F.~F.~Tafti}
\email{Fazel.Fallah.Tafti@USherbrooke.ca}
\affiliation{D\'epartement de physique \& RQMP, Universit\'e de Sherbrooke, Sherbrooke, QC, Canada}

 \author{J. P. Clancy}%
\affiliation{Department of Physics, University of Toronto, Toronto, ON, Canada}%

\author{M. Lapointe-Major}
\affiliation{D\'epartement de physique \& RQMP, Universit\'e de Sherbrooke, Sherbrooke, QC, Canada}

\author{C. Collignon}
\affiliation{D\'epartement de physique \& RQMP, Universit\'e de Sherbrooke, Sherbrooke, QC, Canada}

\author{S. Faucher}
\affiliation{D\'epartement de physique \& RQMP, Universit\'e de Sherbrooke, Sherbrooke, QC, Canada}

\author{J. A. Sears}%
\affiliation{Department of Physics, University of Toronto, Toronto, ON, Canada}%

\author{A. Juneau-Fecteau}
\affiliation{D\'epartement de physique \& RQMP, Universit\'e de Sherbrooke, Sherbrooke, QC, Canada}

\author{N.~Doiron-Leyraud}
\affiliation{D\'epartement de physique \& RQMP, Universit\'e de Sherbrooke, Sherbrooke, QC, Canada}

\author{A. F. Wang}%
\affiliation{Hefei National Laboratory for Physical Sciences at Microscale and Department of Physics, University of Science and Technology of China, Hefei, Anhui 230026, China}%

\author{X.-G.~Luo}%
\affiliation{Hefei National Laboratory for Physical Sciences at Microscale and Department of Physics, University of Science and Technology of China, Hefei, Anhui 230026, China}%

\author{X.~H.~Chen}%
\affiliation{Hefei National Laboratory for Physical Sciences at Microscale and Department of Physics, University of Science and Technology of China, Hefei, Anhui 230026, China}%

\author{S. Desgreniers}%
\affiliation{Faculty of Science, University of Ottawa, Ottawa, ON, Canada}%

\author{Young-June Kim}%
\affiliation{Department of Physics, University of Toronto, Toronto, ON, Canada}%

\author{Louis Taillefer}
\email{Louis.Taillefer@USherbrooke.ca}
\affiliation{D\'epartement de physique \& RQMP, Universit\'e de Sherbrooke, Sherbrooke, QC, Canada}
\affiliation{Canadian Institute for Advanced Research, Toronto, ON, Canada}

\date{\today}

\begin{abstract}
We report a sudden reversal in the pressure dependence of \Tc~in the iron-based superconductor \Cs,
similar to that discovered recently in \K~[Tafti {\it et al.}, Nat. Phys. {\bf 9}, 349 (2013)].
As in \K, we observe no change in the Hall coefficient at $T \to 0$, again ruling out a Lifshitz transition across the critical pressure \Pc.
We interpret the \Tc~reversal in the two materials as a phase transition from one pairing state to another, tuned by pressure,
and we investigate what parameters control this transition.
Comparing samples of different residual resistivity $\rho_0$, we find that a 6-fold increase in impurity scattering does not shift \Pc.
From a study of X-ray diffraction on \K~under pressure, we report the pressure dependence of lattice constants
and As-Fe-As bond angle.
The pressure dependence of the various lattice parameters suggests that \Pc~should be significantly higher in \Cs~than in \K,
but we find on the contrary that \Pc~is lower in \Cs, indicating that other factors control \Tc.
Resistivity measurements under pressure reveal a change of regime across \Pc, suggesting a possible link 
between inelastic scattering and pairing symmetry. 
%
\end{abstract}

\maketitle


\section{\label{Introduction}Introduction}

To understand what controls \Tc~in high temperature superconductors remains a major challenge. 
Several studies suggest that in contrast to cuprates where chemical substitution controls electron concentration, 
the dominant effect of chemical substitution in iron-based superconductors is to tune the structural parameters -- such as the As-Fe-As bond angle -- which in turn control \Tc. \cite{zhao_structural_2008, rotter_superconductivity_2008}
This idea is supported by the parallel tuning of $T_c$ and the structural parameters of the 122 parent compounds BaFe$_2$As$_2$ and SrFe$_2$As$_2$. \cite{kimber_similarities_2009, alireza_superconductivity_2009} 
In the case of Ba$_{1-x}$K$_{x}$Fe$_2$As$_2$, 
at optimal doping ($x=0.4$, \Tc~=~38~K) the As-Fe-As bond angle is $\alpha=109.5\,^{\circ}$, the ideal angle of a non-distorted FeAs$_4$ tetrahedral coordination.
Underdoping, overdoping, or pressure would tune the bond angle away from this ideal value and reduce \Tc~by changing the electronic bandwidth and the nesting conditions. \cite{kimber_similarities_2009} 
\Cs~is an iron-based superconductor with \Tc~=~1.8~K and \Hc~=~1.4~T. \cite{sasmal_superconducting_2008, wang_calorimetric_2013, hong_nodal_2013}
Based on the available X-ray data, \cite{sasmal_superconducting_2008} the As-Fe-As bond angle in \Cs~is $109.58\,^{\circ}$, 
close to the ideal bond angle that yields \Tc~=~38~K in optimally-doped Ba$_{0.6}$K$_{0.4}$Fe$_2$As$_2$.
If the bond angle were the key tuning factor for \Tc, \Cs~should have  a much higher transition temperature than 1.8 K.
In this article, we show evidence that \Tc~in (K,Cs)Fe$_2$As$_2$ may be controlled by details of the inelastic scattering processes 
that are not directly related to structural parameters, but are encoded in the electrical resistivity $\rho(T)$.
The importance of inter- and intra-band inelastic scattering processes in determining \Tc~and the pairing symmetry of iron pnictides 
has been emphasized in several theoretical works. \cite{graser_near-degeneracy_2009, maiti_evolution_2011, maiti_gap_2012} 
Recently, it was shown that a change of pairing symmetry can be induced by tuning the relative strength of different competing inelastic scattering processes,
\ie~different magnetic fluctuation wavevectors. \cite{fernandes_suppression_2013}
%

In a previous paper, we reported the discovery of a sharp reversal in the pressure dependence of \Tc~in \K, the fully hole-doped member of the 
Ba$_{1-x}$K$_x$Fe$_2$As$_2$ series. \cite{tafti_sudden_2013}
No sudden change was observed in the Hall coefficient or resistivity across the critical pressure \Pc~=~17.5~kbar,
indicating that the transition is not triggered by a change in the Fermi surface.
Recent dHvA experiments under pressure confirm that the Fermi surface is the same on both sides of \Pc,
ruling out a Lifshitz transition and strengthening the case for a change of pairing state. \cite{terashima_two_2014}
We interpret the sharp \Tc~reversal as a phase transition from $d$-wave to $s$-wave symmetry.
Bulk measurements such as thermal conductivity\cite{reid_universal_2012, dong_quantum_2010} and penetration depth\cite{hashimoto_evidence_2010} 
favor $d$-wave symmetry at zero pressure.
Because the high-pressure phase is very sensitive to disorder, a likely $s$-wave state is one that changes sign around the Fermi surface,
as in the $s_{\pm}$ state that changes sign between the $\Gamma$-centered hole pockets, as proposed by Maiti \etal \cite{maiti_gap_2012}
It appears that in \K~$s$-wave and $d$-wave states are nearly degenerate, and a small pressure is enough the push the system from one 
state to the other. 
%

In this article, we report the discovery of a similar \Tc~reversal in \Cs.
%
%
The two systems have the same tetragonal structure, but their lattice parameters are notably different. \cite{sasmal_superconducting_2008}
%
%
Our high-pressure X-ray data reveal that at least 30~kbar of pressure is required for the lattice parameters of \Cs~to match those of \K.
%
%
Yet, surprisingly, we find that \Pc~is {\it smaller} in \Cs~than in \K.
This observation clearly shows that structural parameters alone are not the controlling factors for \Pc~in (K,Cs)Fe$_2$As$_2$.
Instead, we propose that competing inelastic scattering processes are responsible for tipping the balance between pairing symmetries.

\begin{figure}
\includegraphics[width=3.5in]{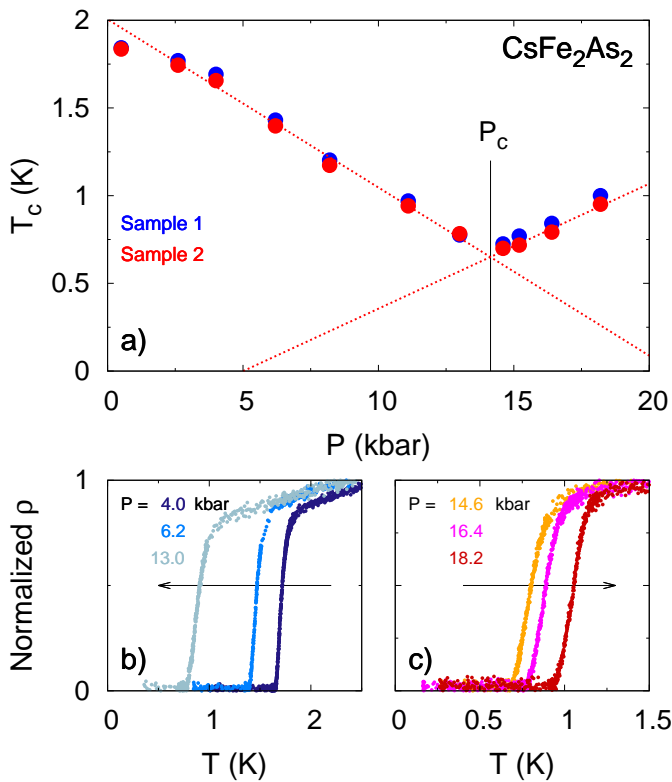}
\caption{\label{resistivity} 
a) Pressure dependence of \Tc~in \Cs. 
The blue and red circles represent data from samples 1 and  2, respectively. 
\Tc~is defined as the temperature where the zero-field resistivity $\rho(T)$ goes to zero.
The critical pressure \Pc~marks a change of behaviour from decreasing to increasing \Tc. 
Dotted red lines are linear fits to the data from sample 2 in the range \Pc~$- 10$ kbar and \Pc~$+ 5$ kbar. 
The critical pressure \Pc~$= 14 \pm 1$~kbar is defined as the intersection of the two linear fits. 
b) Low-temperature $\rho(T)$ data, from sample 2, normalized to unity at $T=2.5$~K. 
Three isobars are shown at $P<$~\Pc, with pressure values as indicated.
The arrow shows that \Tc~decreases with increasing pressure. 
c) Same as in b), but for $P>$~\Pc, with $\rho$ normalized to unity at $T=1.5$~K. 
The arrow shows that \Tc~now {\it increases} with increasing pressure. }
\end{figure}

\section{\label{Experiments}Experiments}

Single crystals of \Cs~were grown using a self-flux method. \cite{hong_nodal_2013}
Resistivity and Hall measurements were performed in in an adiabatic demagnetization refrigerator,
on samples placed inside a clamp cell, using a six-contact configuration.
Hall voltage is measured at plus and minus 10~T from $T=20$ to 0.2~K and antisymmetrized to calculate the Hall coefficient \RH. 
Pressures up to 20~kbar were applied and measured with a precision of $\pm~0.1$ kbar 
by monitoring the superconducting transition temperature of a lead gauge placed besides the samples inside the clamp cell.
A pentane mixture was used as the pressure medium.
%
%
Two samples of \Cs, labelled ``sample 1" and ``sample 2", were measured and excellent reproducibility was observed.
%
%

High pressure X-ray experiments were performed on polycrystalline powder specimens of \K~up to 60~kbar with the HXMA beam line at the Canadian Light Source, using a diamond anvil cell with silicon oil as the pressure medium.
Pressure was tuned blue with a precision of $\pm~2$ kbar using the R$_1$ fluorescent line of a ruby chip placed inside the sample space.
XRD data were collected using angle-dispersive techniques, employing high energy X-rays ($E_i = 24.35$~keV) and a Mar345 image plate detector.
Structural parameters were extracted from full profile Rietveld refinement using the GSAS software. \cite{GSAS_2000} 
Representative refinements of the X-ray data are presented in appendix \ref{rawX}.

\section{\label{Results}Results}

Fig.~\ref{resistivity}a shows our discovery of a sudden reversal in the pressure dependence of \Tc~in \Cs~at a critical pressure \Pc~=~$14\pm1$~kbar.
%
%
The shift of \Tc~as a function of pressure clearly changes direction from decreasing (Fig.~\ref{resistivity}b) to increasing (Fig.~\ref{resistivity}c) across the critical pressure \Pc.
\Tc~varies linearly near \Pc, resulting in a $V$-shaped phase diagram similar to that of \K. \cite{tafti_sudden_2013}

\begin{figure}
\includegraphics[width=3.5in]{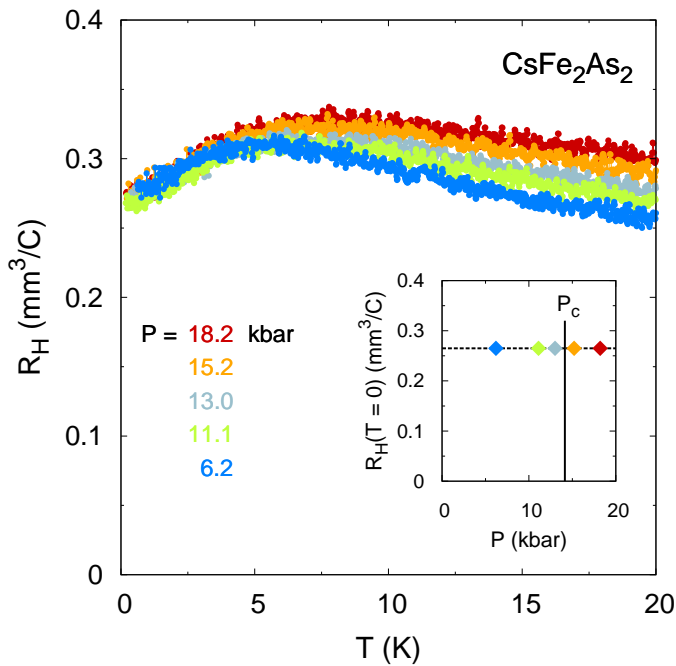}
\caption{\label{hall} 
Temperature dependence of the Hall coefficient \RH$(T)$ in \Cs~(sample 2), at five selected pressures, as indicated. 
The low-temperature data converge to the same value for all pressures, whether below or above \Pc.
{\it Inset}:
The value of \RH~extrapolated to $T=0$ is plotted at different pressures. 
Horizontal and vertical error bars are smaller than symbol dimensions.
\RH$(T=0)$ is seen to remain unchanged across \Pc.
%
}
\end{figure}

\begin{figure}
\includegraphics[width=3.5in]{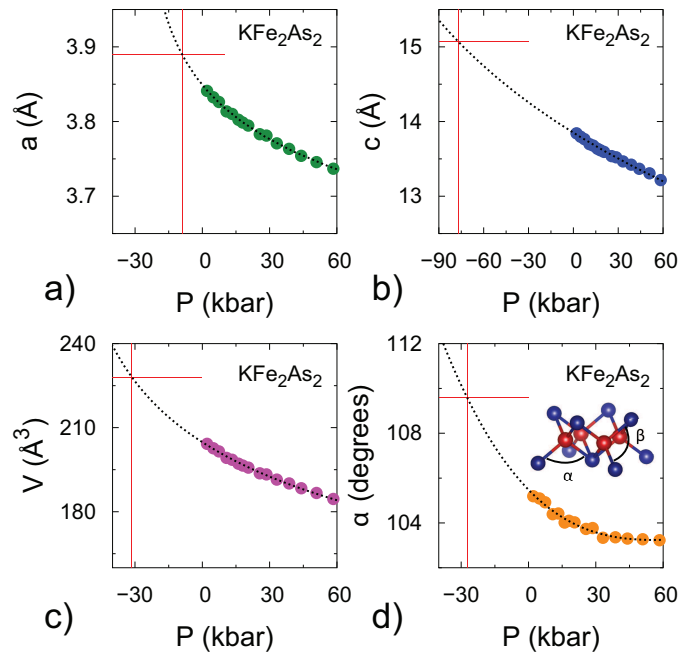}
\caption{\label{lattice} 
Structural parameters of \K~as a function of pressure, up to 60~kbar: 
a) lattice constant $a$;
b) lattice constant $c$;
c) unit cell volume $V=a^2c$;
d) the intra-planar As-Fe-As bond angle $\alpha$ as defined in the {\it inset} (See appendix \ref{alphabeta} for the inter-planar bond angle). 
%
%
Experimental errors on lattice parameters are smaller than symbol dimensions.
The black dotted line in panel a, b, and c is a fit to the standard Murnaghan equation of state extended smoothly to negative pressures. \cite{murnaghan_finite_1937}
From the fits, we extract the moduli of elasticity and report them in appendix \ref{compressibility}.
The black dotted line in panel d is a third order power law fit.
In each panel, the horizontal red line marks the lattice parameter of \Cs,
and the vertical red line gives the negative pressure required for the lattice parameter of \K~to reach the value in \Cs.}
\end{figure}

Measurements of the Hall coefficient \RH~allow us to rule out the possibility of a Lifshitz transition,
\ie~a sudden change in the Fermi surface topology.
%
%
Fig.~\ref{hall} shows the temperature dependence of \RH~at five different pressures.
%
%
In the zero-temperature limit, $R_{\rm H}(T\to 0)$ is seen to remain unchanged across \Pc~(Fig. \ref{hall}, inset).
If the Fermi surface underwent a change, such as the disappearance of one sheet, this would affect $R_{\rm H}(T\to 0)$,
which is a weighted average of the Hall response of the various sheets.
%
%
%
%
Similar Hall measurements were also used to rule out a Lifshitz transition in \K,\cite{tafti_sudden_2013}
in agreement with the lack of any change in dHvA frequencies.\cite{terashima_two_2014}
Several studies on the Ba$_{1-x}$K$_{x}$Fe$_2$As$_2$ series suggest that lattice parameters, in particular the As-Fe-As bond angle, control \Tc. \cite{rotter_superconductivity_2008, kimber_similarities_2009, alireza_superconductivity_2009, budko_heat_2013}
To explore this hypothesis, we measured the lattice parameters of \K~as a function of pressure, up to 60 kbar, 
in order to find out how much pressure is required to tune the lattice parameters of \Cs~so they match those of \K.
Cs has a larger atomic size than K, hence one can view  \Cs~as a negative-pressure version of \K.
The four panels of Fig.~\ref{lattice} show the pressure variation of the lattice constants $a$ and $c$, the unit cell volume ($V=a^2c$), and the intra-planar As-Fe-As bond angle ($\alpha$) in \K.
%
%
The red horizontal line in each panel marks the value of the corresponding lattice parameter in \Cs. \cite{sasmal_superconducting_2008}
In order to tune $a$, $c$, $V$, and $\alpha$ in \K~to match the corresponding values in \Cs, 
a negative pressure of approximately $-10$, $-75$, $-30$, and $-30$ kbar is required, respectively. 
Adding these numbers to the critical pressure for \K~(\Pc~=~17.5~kbar), we would naively estimate that the critical pressure in 
\Cs~should be \Pc~$\simeq 30$~kbar or higher.
We find instead that \Pc~=~14~kbar, 
showing that other factors are involved in controlling \Pc.
%

\begin{figure}
\includegraphics[width=3.5in]{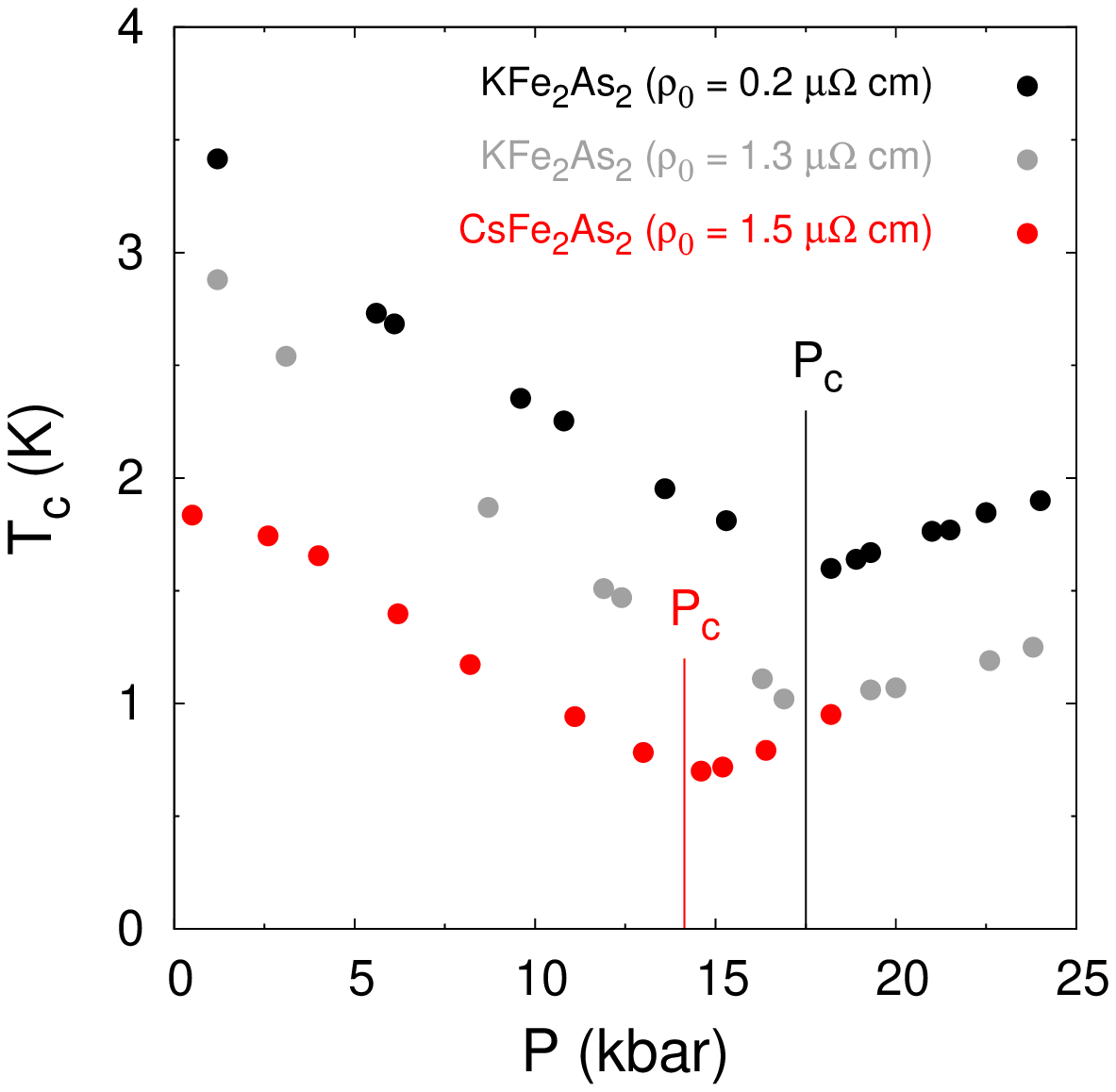}
\caption{\label{PhaseDiagram} 
Pressure dependence of \Tc~in three samples: 
pure \K~(black circles),
less pure \K~(grey circles), 
and \Cs~(sample 2, red circles). 
Even though the \Tc~values for the two \K~samples are different due to different disorder levels, 
measured by their different residual resistivity $\rho_0$,
the critical pressure is the same (\Pc~=~17.5~kbar). 
This shows that the effect of disorder on \Pc~in \K~is negligible.
For comparable $\rho_0$, the critical pressure in \Cs, \Pc~=~14~kbar, is clearly smaller than in \K.}
\end{figure}

\begin{figure}[t]
\includegraphics[width=3.5in]{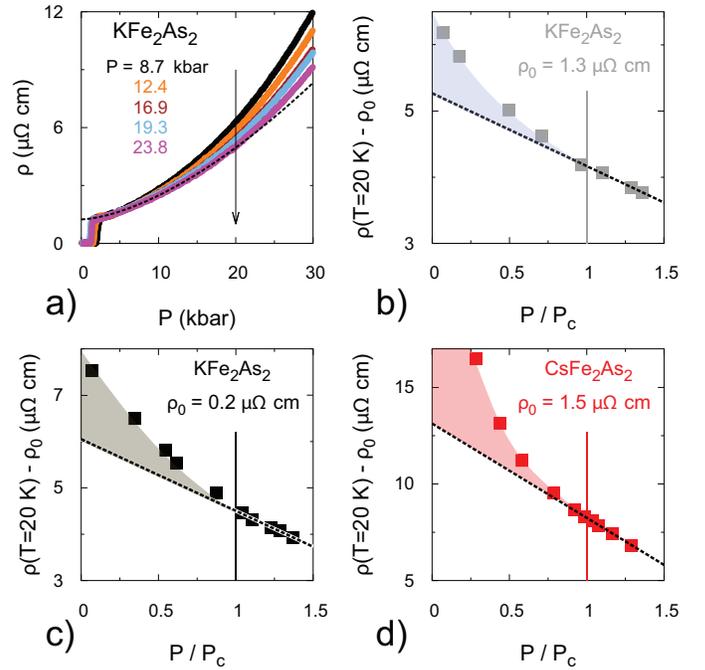}
\caption{\label{InelasticScattering}
a) Resistivity data for the \K~sample with $\rho_0=1.3$~\mucm~at five selected pressures.
The black vertical arrow shows a cut through each curve at $T=20$~K and the dashed line is a power law fit to the curve at $P=23.8$~kbar from 5 to 15~K that is used to extract the residual resistivity $\rho_0$.
Inelastic resistivity, defined as $\rho(T = 20~\rm{K}) - \rho_0$ is plotted vs $P / P_{\rm c}$ in
b) the less pure \K~sample, 
c) the purer \K~sample, and
d) \Cs~(sample 2) 
where \Pc~=~17.5~kbar for \K~and \Pc~=~14~kbar for \Cs.
%
In panel (b), (c), and (d) the dashed black line is a linear fit to the data above  $P / P_{\rm c} = 1$. 
%
}
\end{figure}

It is possible that the lower \Pc~in \Cs~could be due to the fact that \Tc~itself is lower than in \K~at zero pressure,
\ie~that the low-pressure phase is weaker in \Cs.
%
%
One hypothesis for the lower \Tc~in \Cs~is a higher level of disorder.
To test this idea, we studied the pressure dependence of \Tc~in a less pure \K~sample.
Fig.~\ref{PhaseDiagram} compares the $T$-$P$ phase diagram in three samples:
1) a high-purity \K~sample, with $\rho_0=0.2~\mu \Omega$~cm (from ref.~\onlinecite{tafti_sudden_2013});
2) a less pure \K~sample, with $\rho_0=1.3~\mu \Omega$~cm, measured here;
3) a \Cs~sample (sample 2), with $\rho_0=1.5~\mu \Omega$~cm.
Different disorder levels in our samples are due to growth conditions, not to deliberate chemical substitution or impurity inclusions.
First, we observe that a 6-fold increase of $\rho_0$ has negligible impact on \Pc~in \K.
Secondly, we observe that \Pc~is 4 kbar smaller in \Cs~than in \K, for samples of comparable $\rho_0$.
These observations rule out the idea that disorder could be responsible for the lower value of \Pc~in \Cs~compared to \K. 
%

\section{\label{Discussion}Discussion}

We have established a common trait in \Cs~and \K:
both systems have a sudden reversal in the pressure dependence of \Tc,
with no change in the underlying Fermi surface.
%
%
%
%
The question is: what controls that transition?
%
Why does the low-pressure superconducting state become unstable against the high-pressure state?
In a recent theoretical work by Fernandes and Millis, it is demonstrated that different pairing interactions in 122 systems can favour different pairing symmetries. \cite{fernandes_suppression_2013}
In their model, SDW-type magnetic fluctuations, with wavevector $(\pi,0)$,  favour  $s_{\pm}$ pairing, whereas
 N\'eel-type fluctuations, with wavevector $(\pi,\pi)$, strongly suppress the $s_{\pm}$ state and favour 
$d$-wave pairing.
A gradual increase in the $(\pi,\pi)$ fluctuations eventually causes a phase transition from an $s_{\pm}$ 
superconducting state to a $d$-wave state, producing a V-shaped \Tc~vs $P$ curve.\cite{fernandes_suppression_2013}
In \K~and \Cs, it is conceivable that two such competing interactions are at play, with pressure tilting the balance in favor of one versus the other.
We explore such a scenario by looking at how the inelastic scattering evolves with pressure, measured via the inelastic
resistivity, defined as $\rho(T) - \rho_0$, where $\rho_0$ is the residual resistivity.
%
Fig.~\ref{InelasticScattering}(a) shows raw resistivity data from the \K~sample with $\rho_0=1.3$~\mucm~below 30~K.
To extract $\rho(T) - \rho_0$ at each pressure, we make a cut through each curve at $T=20$~K and subtract from it the residual resistivity $\rho_0$ that comes from a power-law fit $\rho=\rho_0+AT^n$ to each curve.
$\rho_0$ is determined by disorder level and does not change as a function of pressure.
The resulting $\rho(T=20~\rm{K})-\rho_0$ values for this sample are then plotted as a function of normalized pressure $P / $\Pc~in Fig.~\ref{InelasticScattering}(b). 
Through a similar process we extract the pressure dependence of $\rho(20~\rm{K})-\rho_0$ in \Cs~and the purer \K~sample with $\rho_0 = 0.2$~\mucm~in Fig.~\ref{InelasticScattering}(c) and (d). 
%
%
%
In all three samples, at $P / $\Pc~$>1$, the inelastic resistivity varies linearly with pressure.
%
As $P$ drops below \Pc,  the inelastic resistivity in (K,Cs)Fe$_2$As$_2$ shows a clear rise below their respective \Pc,
over and above the linear regime.
Fig. \ref{InelasticScattering} therefore suggests a connection between the transition in the pressure dependence of \Tc~
and the appearance of an additional inelastic scattering process.
Note that our choice of $T=20$~K for the inelastic resistivity is arbitrary.
Resistivity cuts at any finite temperature above $T_c$ give qualitatively similar results. 

The Fermi surface of \K~includes three $\Gamma$-centered hole-like cylinders.
A possible pairing state is an $s_{\pm}$ state where the change of sign occurs between the inner cylinder and
the middle cylinder, favored by a small-$Q$ interaction. \cite{maiti_gap_2012}
By contrast, the intraband inelastic scattering wavevectors that favour $d$-wave pairing
are large-$Q$ processes. \cite{thomale_exotic_2011}
Therefore, one scenario in which to understand the evolution in the inelastic resistivity with pressure (Fig.~5), 
and its link to the \Tc~reversal, is the following.
At low pressure, the large-$Q$ scattering processes that favor $d$-wave pairing make a substantial contribution to the resistivity,
as they produce a large change in momentum.
These weaken with pressure, causing a decrease in both \Tc~and the resistivity.
This decrease persists  
until the low-$Q$ processes that favor  $s_{\pm}$ pairing, less visible in the resistivity, come to dominate,
above \Pc. 
%
%

%

In summary, we discovered a pressure-induced reversal in the dependence of the transition temperature
 \Tc~on pressure in the iron-based superconductor \Cs, similar to a our previous finding in \K.
We interpret the \Tc~reversal at the critical pressure \Pc~as a transition from one pairing state to another.
The fact that \Pc~in \Cs~is smaller than in \K,
even though all lattice parameters would suggest otherwise,
shows that structural parameters alone do not control \Pc.
We also demonstrate that disorder has negligible effect on \Pc.
Our study of the pressure dependence of resistivity in \Cs~and \K~reveals a possible link between \Tc~and inelastic scattering.
Our proposal is that the high-pressure phase in both materials is an $s_{\pm}$ state that changes sign between $\Gamma$-centered pockets.
As the pressure is lowered, the large-$Q$ inelastic scattering processes that favor $d$-wave pairing in pure  \K~and \Cs~grow until
at a critical pressure \Pc~they cause a transition from one superconducting state to another, with a change of pairing symmetry from $s$-wave to $d$-wave.
The experimental evidence for this is the fact that below \Pc~the inelastic resistivity, measured as the difference $\rho(20~\mathrm{K})-\rho_0$, 
deviates upwards from its linear pressure dependence at high pressure.
%


\section*{ACKNOWLEDGMENTS}
We thank 
A.~V.~Chubukov, R.~M.~Fernandes and A.~J.~Millis for helpful discussions, and
S.~Fortier for his assistance with the experiments. 
The work at Sherbrooke was supported by the Canadian Institute for Advanced Research and a Canada Research Chair and it was funded by NSERC,
FRQNT and CFI. 
Work done in China was supported by the National Natural Science Foundation of China (Grant No. 11190021), 
the Strategic Priority Research Program (B) of the Chinese Academy of Sciences, 
and the National Basic Research Program of China.
Research at the University of Toronto was supported by the NSERC, CFI, Onatrio Ministry of Research and Innovation, and Canada Research Chair program.
The Canadian Light Source is funded by CFI, NSERC, the National Research Council Canada, the Canadian Institutes
of Health Research, the Government of Saskatchewan, Western Economic Diversification Canada, and the University
of Saskatchewan.

\appendix

\section{\label{rawX}X-ray data}

\begin{figure}[t]
\includegraphics[width=3.5in]{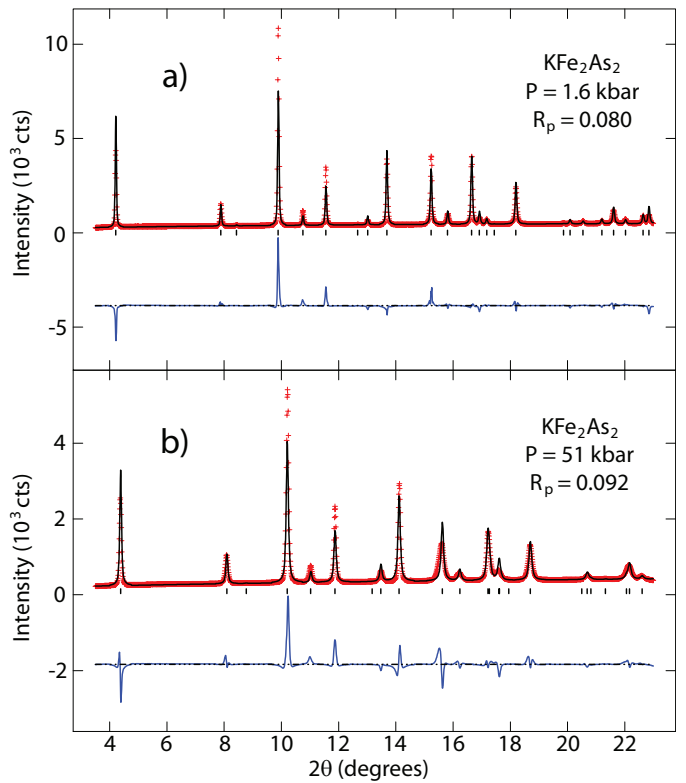}
\caption{\label{Representative_Refinements}
Representative refinement of the X-ray diffraction patterns collected at a) $P=1.6$~kbar and b) $P=51$~kbar.
Red crosses show the XRD data plotted as intensity versus $2\Theta$.
Black lines are the best fit to the data.
Blue lines show the difference between the data and the fits.
The goodness of the fit parameter ($\rm{R_p}$) is provided for each refinement.
}
\end{figure}

All our X-ray measurements are performed at room temperature using angle-dispersive technique with the HXMA beam line at CLS. 
Figure~\ref{Representative_Refinements} includes two representative structural refinements of the X-ray diffraction data at $P=1.6$~kbar and $P=51$~kbar.
2D diffraction data from the image plate detector were reduced to 1D using the FIT2D program \cite{hammersley_two-dimensional_1996} and plotted as intensity vs $2\Theta$.
The structural refinements were performed using the GSAS software package. \cite{GSAS_2000}
%
%
The experimental data points are illustrated by red crosses, 
the best fit to the diffraction pattern is illustrated by the solid black line, 
and the difference between the two curves is denoted by the solid blue line.  
The Bragg reflections corresponding to the tetragonal $I4/mmm$ structure of \K~are indicated by the black tick marks below
the data.

\section{\label{alphabeta}Bond angles}

\begin{figure}[t]
\includegraphics[width=3.5in]{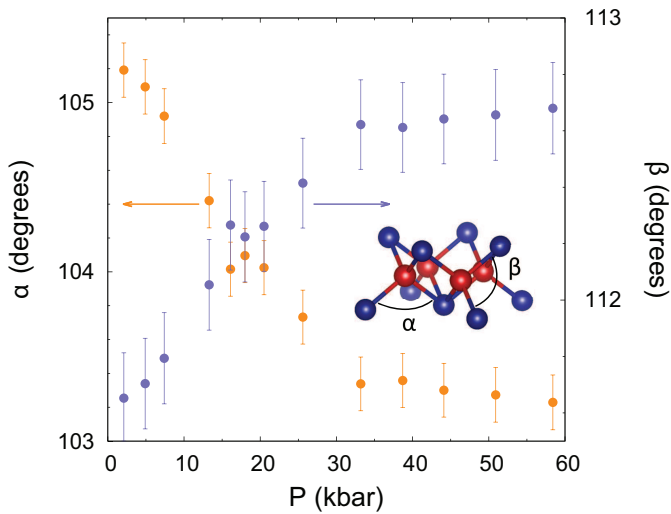}
\caption{\label{BondAngles}
Pressure dependence of both intra-planar ($\alpha$) and inter-planar ($\beta$) bond angles from 0 to 60 kbar.
The values for the two bond angles -- defined in the {\it inset} -- are extracted from structural refinements performed on the X-ray data.
$\alpha$ decreases as a function of pressure while $\beta$ increases.
}
\end{figure}

Within the tetragonal structure of \K, there are two bond angles in each FeAs$_4$ tetrahedron \cite{jeffries_suppression_2012} as indicated in the {\it inset} of Fig.~\ref{BondAngles}: 
The {\it intra-planar} bond angle ($\alpha$) that spans over the bond from one As plane to an Fe atom and back to an As atom in the original plane and the {\it inter-planar} bond angle ($\beta$) that spans over the bond from one As plane through an Fe atom to the next As plane. 
In the case of an ideal undistorted tetrahedron $\alpha = \beta = 109.47^{\circ}$.
In Fig.~\ref{lattice}(d) we present only the intra-planar bond angle $\alpha$ to show that about $-30$ kbar is required to tune $\alpha$ from its value in \K~to \Cs. 
For completeness, here we plot the pressure evolution of both bond angles in  Fig.~\ref{BondAngles}. 
$\alpha$ decreases as a function of pressure while $\beta$ increases, hence, the size of the tetragonal distortion in  
\K~grows progressively larger as the pressure increases.
Interestingly, the form of this tetragonal distortion is opposite to that observed in Ca$_{0.67}$Sr$_{0.33}$Fe$_2$As$_2$ where applied pressure causes intra-layer bond angles to increase and inter-layer bond angles to decrease. \cite{jeffries_suppression_2012}

\section{\label{compressibility}Anisotropic compressibility in \K}

\begin{table}[b]
\caption{\label{bulkmodulus} 
The moduli of elasticity along $a$-axis $K_a$ and $c$-axis $K_c$ as well as the bulk modulus $K$ are extracted by fitting our data to the Murnaghan equation of state. The pressure derivatives of $K_a$, $K_c$, and $K_V$ are also reported. 
}
\begin{ruledtabular}
\begin{tabular}{cccccc}
$K_a \textrm{(GPa)}$&
$K_c \textrm{(GPa)}$&
$K \textrm{(GPa)}$&
$K'_a$&
$K'_c$&
$K'$
\\
\colrule 
\\

105~$\pm$~5  & 115~$\pm$~3  & 40~$\pm$~1  & 400~$\pm$~2 & 3.3~$\pm$~0.8  & 6.1~$\pm$~0.4   \\

\end{tabular}
\end{ruledtabular}
\end{table}

In Fig.~\ref{lattice}, we fit our data to the Murnaghan equation of state: \cite{murnaghan_finite_1937} 
\begin{equation}
P(V) = \frac{K}{K'}\left[ \left( \frac{V}{V_0} \right)^{-K'}-1 \right]
\label{Murnaghan}
\end{equation}
and extend it smoothly to negative pressures to find how much pressure is required to tune the lattice parameters of \K~to those of \Cs.
Note that the compressibility of \K~appears to be anisotropic.
The fits also allow us to extract the bulk modulus $K$ and its pressure derivative $K'=\partial K/ \partial P$ in \K. 
Table~\ref{bulkmodulus} summarizes the values of the bulk modulus $K$ as well as the moduli of elasticity along the $a$- and $c$-axes.
%
The modulus of elasticity appears to be almost identical along the $a$- and the $c$-axes, but the first derivative of the modulus is over an order of magnitude larger along the $a$-axis.
This accounts for the roughly 40\% smaller compression observed for the in-plane lattice constant.


\bibliography{CsFe2As2_25mar2014}

\begin{thebibliography}{22}%
\makeatletter
\providecommand \@ifxundefined [1]{%
 \@ifx{#1\undefined}
}%
\providecommand \@ifnum [1]{%
 \ifnum #1\expandafter \@firstoftwo
 \else \expandafter \@secondoftwo
 \fi
}%
\providecommand \@ifx [1]{%
 \ifx #1\expandafter \@firstoftwo
 \else \expandafter \@secondoftwo
 \fi
}%
\providecommand \natexlab [1]{#1}%
\providecommand \enquote  [1]{``#1''}%
\providecommand \bibnamefont  [1]{#1}%
\providecommand \bibfnamefont [1]{#1}%
\providecommand \citenamefont [1]{#1}%
\providecommand \href@noop [0]{\@secondoftwo}%
\providecommand \href [0]{\begingroup \@sanitize@url \@href}%
\providecommand \@href[1]{\@@startlink{#1}\@@href}%
\providecommand \@@href[1]{\endgroup#1\@@endlink}%
\providecommand \@sanitize@url [0]{\catcode `\\12\catcode `\$12\catcode
  `\&12\catcode `\#12\catcode `\^12\catcode `\_12\catcode `\%12\relax}%
\providecommand \@@startlink[1]{}%
\providecommand \@@endlink[0]{}%
\providecommand \url  [0]{\begingroup\@sanitize@url \@url }%
\providecommand \@url [1]{\endgroup\@href {#1}{\urlprefix }}%
\providecommand \urlprefix  [0]{URL }%
\providecommand \Eprint [0]{\href }%
\providecommand \doibase [0]{http://dx.doi.org/}%
\providecommand \selectlanguage [0]{\@gobble}%
\providecommand \bibinfo  [0]{\@secondoftwo}%
\providecommand \bibfield  [0]{\@secondoftwo}%
\providecommand \translation [1]{[#1]}%
\providecommand \BibitemOpen [0]{}%
\providecommand \bibitemStop [0]{}%
\providecommand \bibitemNoStop [0]{.\EOS\space}%
\providecommand \EOS [0]{\spacefactor3000\relax}%
\providecommand \BibitemShut  [1]{\csname bibitem#1\endcsname}%
\let\auto@bib@innerbib\@empty
\bibitem [{\citenamefont {Zhao}\ \emph {et~al.}(2008)\citenamefont {Zhao},
  \citenamefont {Huang}, \citenamefont {de~la Cruz}, \citenamefont {Li},
  \citenamefont {Lynn}, \citenamefont {Chen}, \citenamefont {Green},
  \citenamefont {Chen}, \citenamefont {Li}, \citenamefont {Li}, \citenamefont
  {Luo}, \citenamefont {Wang},\ and\ \citenamefont
  {Dai}}]{zhao_structural_2008}%
  \BibitemOpen
  \bibfield  {author} {\bibinfo {author} {\bibfnamefont {J.}~\bibnamefont
  {Zhao}}, \bibinfo {author} {\bibfnamefont {Q.}~\bibnamefont {Huang}},
  \bibinfo {author} {\bibfnamefont {C.}~\bibnamefont {de~la Cruz}}, \bibinfo
  {author} {\bibfnamefont {S.}~\bibnamefont {Li}}, \bibinfo {author}
  {\bibfnamefont {J.~W.}\ \bibnamefont {Lynn}}, \bibinfo {author}
  {\bibfnamefont {Y.}~\bibnamefont {Chen}}, \bibinfo {author} {\bibfnamefont
  {M.~A.}\ \bibnamefont {Green}}, \bibinfo {author} {\bibfnamefont {G.~F.}\
  \bibnamefont {Chen}}, \bibinfo {author} {\bibfnamefont {G.}~\bibnamefont
  {Li}}, \bibinfo {author} {\bibfnamefont {Z.}~\bibnamefont {Li}}, \bibinfo
  {author} {\bibfnamefont {J.~L.}\ \bibnamefont {Luo}}, \bibinfo {author}
  {\bibfnamefont {N.~L.}\ \bibnamefont {Wang}}, \ and\ \bibinfo {author}
  {\bibfnamefont {P.}~\bibnamefont {Dai}},\ }\href {\doibase 10.1038/nmat2315}
  {\bibfield  {journal} {\bibinfo  {journal} {Nature Materials}\ }\textbf
  {\bibinfo {volume} {7}},\ \bibinfo {pages} {953} (\bibinfo {year}
  {2008})}\BibitemShut {NoStop}%
\bibitem [{\citenamefont {Rotter}\ \emph {et~al.}(2008)\citenamefont {Rotter},
  \citenamefont {Pangerl}, \citenamefont {Tegel},\ and\ \citenamefont
  {Johrendt}}]{rotter_superconductivity_2008}%
  \BibitemOpen
  \bibfield  {author} {\bibinfo {author} {\bibfnamefont {M.}~\bibnamefont
  {Rotter}}, \bibinfo {author} {\bibfnamefont {M.}~\bibnamefont {Pangerl}},
  \bibinfo {author} {\bibfnamefont {M.}~\bibnamefont {Tegel}}, \ and\ \bibinfo
  {author} {\bibfnamefont {D.}~\bibnamefont {Johrendt}},\ }\href {\doibase
  10.1002/anie.200803641} {\bibfield  {journal} {\bibinfo  {journal}
  {Angewandte Chemie International Edition}\ }\textbf {\bibinfo {volume}
  {47}},\ \bibinfo {pages} {7949–7952} (\bibinfo {year} {2008})}\BibitemShut
  {NoStop}%
\bibitem [{\citenamefont {Kimber}\ \emph {et~al.}(2009)\citenamefont {Kimber},
  \citenamefont {Kreyssig}, \citenamefont {Zhang}, \citenamefont {Jeschke},
  \citenamefont {Valentí}, \citenamefont {Yokaichiya}, \citenamefont
  {Colombier}, \citenamefont {Yan}, \citenamefont {Hansen}, \citenamefont
  {Chatterji}, \citenamefont {{McQueeney}}, \citenamefont {Canfield},
  \citenamefont {Goldman},\ and\ \citenamefont
  {Argyriou}}]{kimber_similarities_2009}%
  \BibitemOpen
  \bibfield  {author} {\bibinfo {author} {\bibfnamefont {S.~A.~J.}\
  \bibnamefont {Kimber}}, \bibinfo {author} {\bibfnamefont {A.}~\bibnamefont
  {Kreyssig}}, \bibinfo {author} {\bibfnamefont {Y.-Z.}\ \bibnamefont {Zhang}},
  \bibinfo {author} {\bibfnamefont {H.~O.}\ \bibnamefont {Jeschke}}, \bibinfo
  {author} {\bibfnamefont {R.}~\bibnamefont {Valentí}}, \bibinfo {author}
  {\bibfnamefont {F.}~\bibnamefont {Yokaichiya}}, \bibinfo {author}
  {\bibfnamefont {E.}~\bibnamefont {Colombier}}, \bibinfo {author}
  {\bibfnamefont {J.}~\bibnamefont {Yan}}, \bibinfo {author} {\bibfnamefont
  {T.~C.}\ \bibnamefont {Hansen}}, \bibinfo {author} {\bibfnamefont
  {T.}~\bibnamefont {Chatterji}}, \bibinfo {author} {\bibfnamefont {R.~J.}\
  \bibnamefont {{McQueeney}}}, \bibinfo {author} {\bibfnamefont {P.~C.}\
  \bibnamefont {Canfield}}, \bibinfo {author} {\bibfnamefont {A.~I.}\
  \bibnamefont {Goldman}}, \ and\ \bibinfo {author} {\bibfnamefont {D.~N.}\
  \bibnamefont {Argyriou}},\ }\href {\doibase 10.1038/nmat2443} {\bibfield
  {journal} {\bibinfo  {journal} {Nature Materials}\ }\textbf {\bibinfo
  {volume} {8}},\ \bibinfo {pages} {471} (\bibinfo {year} {2009})}\BibitemShut
  {NoStop}%
\bibitem [{\citenamefont {Alireza}\ \emph {et~al.}(2009)\citenamefont
  {Alireza}, \citenamefont {Ko}, \citenamefont {Gillett}, \citenamefont
  {Petrone}, \citenamefont {Cole}, \citenamefont {Lonzarich},\ and\
  \citenamefont {Sebastian}}]{alireza_superconductivity_2009}%
  \BibitemOpen
  \bibfield  {author} {\bibinfo {author} {\bibfnamefont {P.~L.}\ \bibnamefont
  {Alireza}}, \bibinfo {author} {\bibfnamefont {Y.~T.~C.}\ \bibnamefont {Ko}},
  \bibinfo {author} {\bibfnamefont {J.}~\bibnamefont {Gillett}}, \bibinfo
  {author} {\bibfnamefont {C.~M.}\ \bibnamefont {Petrone}}, \bibinfo {author}
  {\bibfnamefont {J.~M.}\ \bibnamefont {Cole}}, \bibinfo {author}
  {\bibfnamefont {G.~G.}\ \bibnamefont {Lonzarich}}, \ and\ \bibinfo {author}
  {\bibfnamefont {S.~E.}\ \bibnamefont {Sebastian}},\ }\href {\doibase
  10.1088/0953-8984/21/1/012208} {\bibfield  {journal} {\bibinfo  {journal}
  {Journal of Physics: Condensed Matter}\ }\textbf {\bibinfo {volume} {21}},\
  \bibinfo {pages} {012208} (\bibinfo {year} {2009})}\BibitemShut {NoStop}%
\bibitem [{\citenamefont {Sasmal}\ \emph {et~al.}(2008)\citenamefont {Sasmal},
  \citenamefont {Lv}, \citenamefont {Lorenz}, \citenamefont {Guloy},
  \citenamefont {Chen}, \citenamefont {Xue},\ and\ \citenamefont
  {Chu}}]{sasmal_superconducting_2008}%
  \BibitemOpen
  \bibfield  {author} {\bibinfo {author} {\bibfnamefont {K.}~\bibnamefont
  {Sasmal}}, \bibinfo {author} {\bibfnamefont {B.}~\bibnamefont {Lv}}, \bibinfo
  {author} {\bibfnamefont {B.}~\bibnamefont {Lorenz}}, \bibinfo {author}
  {\bibfnamefont {A.~M.}\ \bibnamefont {Guloy}}, \bibinfo {author}
  {\bibfnamefont {F.}~\bibnamefont {Chen}}, \bibinfo {author} {\bibfnamefont
  {Y.-Y.}\ \bibnamefont {Xue}}, \ and\ \bibinfo {author} {\bibfnamefont
  {C.-W.}\ \bibnamefont {Chu}},\ }\href {\doibase
  10.1103/PhysRevLett.101.107007} {\bibfield  {journal} {\bibinfo  {journal}
  {Physical Review Letters}\ }\textbf {\bibinfo {volume} {101}},\ \bibinfo
  {pages} {107007} (\bibinfo {year} {2008})}\BibitemShut {NoStop}%
\bibitem [{\citenamefont {Wang}\ \emph {et~al.}(2013)\citenamefont {Wang},
  \citenamefont {Pan}, \citenamefont {Luo}, \citenamefont {Chen}, \citenamefont
  {Yan}, \citenamefont {Ying}, \citenamefont {Ye}, \citenamefont {Cheng},
  \citenamefont {Hong}, \citenamefont {Li},\ and\ \citenamefont
  {Chen}}]{wang_calorimetric_2013}%
  \BibitemOpen
  \bibfield  {author} {\bibinfo {author} {\bibfnamefont {A.~F.}\ \bibnamefont
  {Wang}}, \bibinfo {author} {\bibfnamefont {B.~Y.}\ \bibnamefont {Pan}},
  \bibinfo {author} {\bibfnamefont {X.~G.}\ \bibnamefont {Luo}}, \bibinfo
  {author} {\bibfnamefont {F.}~\bibnamefont {Chen}}, \bibinfo {author}
  {\bibfnamefont {Y.~J.}\ \bibnamefont {Yan}}, \bibinfo {author} {\bibfnamefont
  {J.~J.}\ \bibnamefont {Ying}}, \bibinfo {author} {\bibfnamefont {G.~J.}\
  \bibnamefont {Ye}}, \bibinfo {author} {\bibfnamefont {P.}~\bibnamefont
  {Cheng}}, \bibinfo {author} {\bibfnamefont {X.~C.}\ \bibnamefont {Hong}},
  \bibinfo {author} {\bibfnamefont {S.~Y.}\ \bibnamefont {Li}}, \ and\ \bibinfo
  {author} {\bibfnamefont {X.~H.}\ \bibnamefont {Chen}},\ }\href {\doibase
  10.1103/PhysRevB.87.214509} {\bibfield  {journal} {\bibinfo  {journal}
  {Physical Review B}\ }\textbf {\bibinfo {volume} {87}},\ \bibinfo {pages}
  {214509} (\bibinfo {year} {2013})}\BibitemShut {NoStop}%
\bibitem [{\citenamefont {Hong}\ \emph {et~al.}(2013)\citenamefont {Hong},
  \citenamefont {Li}, \citenamefont {Pan}, \citenamefont {He}, \citenamefont
  {Wang}, \citenamefont {Luo}, \citenamefont {Chen},\ and\ \citenamefont
  {Li}}]{hong_nodal_2013}%
  \BibitemOpen
  \bibfield  {author} {\bibinfo {author} {\bibfnamefont {X.~C.}\ \bibnamefont
  {Hong}}, \bibinfo {author} {\bibfnamefont {X.~L.}\ \bibnamefont {Li}},
  \bibinfo {author} {\bibfnamefont {B.~Y.}\ \bibnamefont {Pan}}, \bibinfo
  {author} {\bibfnamefont {L.~P.}\ \bibnamefont {He}}, \bibinfo {author}
  {\bibfnamefont {A.~F.}\ \bibnamefont {Wang}}, \bibinfo {author}
  {\bibfnamefont {X.~G.}\ \bibnamefont {Luo}}, \bibinfo {author} {\bibfnamefont
  {X.~H.}\ \bibnamefont {Chen}}, \ and\ \bibinfo {author} {\bibfnamefont
  {S.~Y.}\ \bibnamefont {Li}},\ }\href {\doibase 10.1103/PhysRevB.87.144502}
  {\bibfield  {journal} {\bibinfo  {journal} {Physical Review B}\ }\textbf
  {\bibinfo {volume} {87}},\ \bibinfo {pages} {144502} (\bibinfo {year}
  {2013})}\BibitemShut {NoStop}%
\bibitem [{\citenamefont {Graser}\ \emph {et~al.}(2009)\citenamefont {Graser},
  \citenamefont {Maier}, \citenamefont {Hirschfeld},\ and\ \citenamefont
  {Scalapino}}]{graser_near-degeneracy_2009}%
  \BibitemOpen
  \bibfield  {author} {\bibinfo {author} {\bibfnamefont {S.}~\bibnamefont
  {Graser}}, \bibinfo {author} {\bibfnamefont {T.~A.}\ \bibnamefont {Maier}},
  \bibinfo {author} {\bibfnamefont {P.~J.}\ \bibnamefont {Hirschfeld}}, \ and\
  \bibinfo {author} {\bibfnamefont {D.~J.}\ \bibnamefont {Scalapino}},\ }\href
  {\doibase 10.1088/1367-2630/11/2/025016} {\bibfield  {journal} {\bibinfo
  {journal} {New Journal of Physics}\ }\textbf {\bibinfo {volume} {11}},\
  \bibinfo {pages} {025016} (\bibinfo {year} {2009})}\BibitemShut {NoStop}%
\bibitem [{\citenamefont {Maiti}\ \emph {et~al.}(2011)\citenamefont {Maiti},
  \citenamefont {Korshunov}, \citenamefont {Maier}, \citenamefont
  {Hirschfeld},\ and\ \citenamefont {Chubukov}}]{maiti_evolution_2011}%
  \BibitemOpen
  \bibfield  {author} {\bibinfo {author} {\bibfnamefont {S.}~\bibnamefont
  {Maiti}}, \bibinfo {author} {\bibfnamefont {M.~M.}\ \bibnamefont
  {Korshunov}}, \bibinfo {author} {\bibfnamefont {T.~A.}\ \bibnamefont
  {Maier}}, \bibinfo {author} {\bibfnamefont {P.~J.}\ \bibnamefont
  {Hirschfeld}}, \ and\ \bibinfo {author} {\bibfnamefont {A.~V.}\ \bibnamefont
  {Chubukov}},\ }\href {\doibase 10.1103/PhysRevLett.107.147002} {\bibfield
  {journal} {\bibinfo  {journal} {Physical Review Letters}\ }\textbf {\bibinfo
  {volume} {107}},\ \bibinfo {pages} {147002} (\bibinfo {year}
  {2011})}\BibitemShut {NoStop}%
\bibitem [{\citenamefont {Maiti}\ \emph {et~al.}(2012)\citenamefont {Maiti},
  \citenamefont {Korshunov},\ and\ \citenamefont {Chubukov}}]{maiti_gap_2012}%
  \BibitemOpen
  \bibfield  {author} {\bibinfo {author} {\bibfnamefont {S.}~\bibnamefont
  {Maiti}}, \bibinfo {author} {\bibfnamefont {M.~M.}\ \bibnamefont
  {Korshunov}}, \ and\ \bibinfo {author} {\bibfnamefont {A.~V.}\ \bibnamefont
  {Chubukov}},\ }\href {\doibase 10.1103/PhysRevB.85.014511} {\bibfield
  {journal} {\bibinfo  {journal} {Physical Review B}\ }\textbf {\bibinfo
  {volume} {85}},\ \bibinfo {pages} {014511} (\bibinfo {year}
  {2012})}\BibitemShut {NoStop}%
\bibitem [{\citenamefont {Fernandes}\ and\ \citenamefont
  {Millis}(2013)}]{fernandes_suppression_2013}%
  \BibitemOpen
  \bibfield  {author} {\bibinfo {author} {\bibfnamefont {R.~M.}\ \bibnamefont
  {Fernandes}}\ and\ \bibinfo {author} {\bibfnamefont {A.~J.}\ \bibnamefont
  {Millis}},\ }\href {\doibase 10.1103/PhysRevLett.110.117004} {\bibfield
  {journal} {\bibinfo  {journal} {Physical Review Letters}\ }\textbf {\bibinfo
  {volume} {110}},\ \bibinfo {pages} {117004} (\bibinfo {year}
  {2013})}\BibitemShut {NoStop}%
\bibitem [{\citenamefont {Tafti}\ \emph {et~al.}(2013)\citenamefont {Tafti},
  \citenamefont {Juneau-Fecteau}, \citenamefont {Delage}, \citenamefont
  {René~de Cotret}, \citenamefont {Reid}, \citenamefont {Wang}, \citenamefont
  {Luo}, \citenamefont {Chen}, \citenamefont {Doiron-Leyraud},\ and\
  \citenamefont {Taillefer}}]{tafti_sudden_2013}%
  \BibitemOpen
  \bibfield  {author} {\bibinfo {author} {\bibfnamefont {F.~F.}\ \bibnamefont
  {Tafti}}, \bibinfo {author} {\bibfnamefont {A.}~\bibnamefont
  {Juneau-Fecteau}}, \bibinfo {author} {\bibfnamefont {M.-.}\ \bibnamefont
  {Delage}}, \bibinfo {author} {\bibfnamefont {S.}~\bibnamefont {René~de
  Cotret}}, \bibinfo {author} {\bibfnamefont {J.-P.}\ \bibnamefont {Reid}},
  \bibinfo {author} {\bibfnamefont {A.~F.}\ \bibnamefont {Wang}}, \bibinfo
  {author} {\bibfnamefont {X.-G.}\ \bibnamefont {Luo}}, \bibinfo {author}
  {\bibfnamefont {X.~H.}\ \bibnamefont {Chen}}, \bibinfo {author}
  {\bibfnamefont {N.}~\bibnamefont {Doiron-Leyraud}}, \ and\ \bibinfo {author}
  {\bibfnamefont {L.}~\bibnamefont {Taillefer}},\ }\href {\doibase
  10.1038/nphys2617} {\bibfield  {journal} {\bibinfo  {journal} {Nature
  Physics}\ }\textbf {\bibinfo {volume} {9}},\ \bibinfo {pages} {349} (\bibinfo
  {year} {2013})}\BibitemShut {NoStop}%
\bibitem [{\citenamefont {Terashima}\ \emph {et~al.}(2014)\citenamefont
  {Terashima}, \citenamefont {Kihou}, \citenamefont {Sugii}, \citenamefont
  {Kikugawa}, \citenamefont {Matsumoto}, \citenamefont {Ishida}, \citenamefont
  {Lee}, \citenamefont {Iyo}, \citenamefont {Eisaki},\ and\ \citenamefont
  {Uji}}]{terashima_two_2014}%
  \BibitemOpen
  \bibfield  {author} {\bibinfo {author} {\bibfnamefont {T.}~\bibnamefont
  {Terashima}}, \bibinfo {author} {\bibfnamefont {K.}~\bibnamefont {Kihou}},
  \bibinfo {author} {\bibfnamefont {K.}~\bibnamefont {Sugii}}, \bibinfo
  {author} {\bibfnamefont {N.}~\bibnamefont {Kikugawa}}, \bibinfo {author}
  {\bibfnamefont {T.}~\bibnamefont {Matsumoto}}, \bibinfo {author}
  {\bibfnamefont {S.}~\bibnamefont {Ishida}}, \bibinfo {author} {\bibfnamefont
  {C.-H.}\ \bibnamefont {Lee}}, \bibinfo {author} {\bibfnamefont
  {A.}~\bibnamefont {Iyo}}, \bibinfo {author} {\bibfnamefont {H.}~\bibnamefont
  {Eisaki}}, \ and\ \bibinfo {author} {\bibfnamefont {S.}~\bibnamefont {Uji}},\
  }\href {http://arxiv.org/abs/1401.6257} {\bibfield  {journal} {\bibinfo
  {journal} {{arXiv:1401.6257} [cond-mat]}\ } (\bibinfo {year}
  {2014})}\BibitemShut {NoStop}%
\bibitem [{\citenamefont {Reid}\ \emph {et~al.}(2012)\citenamefont {Reid},
  \citenamefont {Tanatar}, \citenamefont {Juneau-Fecteau}, \citenamefont
  {Gordon}, \citenamefont {de~Cotret}, \citenamefont {Doiron-Leyraud},
  \citenamefont {Saito}, \citenamefont {Fukazawa}, \citenamefont {Kohori},
  \citenamefont {Kihou}, \citenamefont {Lee}, \citenamefont {Iyo},
  \citenamefont {Eisaki}, \citenamefont {Prozorov},\ and\ \citenamefont
  {Taillefer}}]{reid_universal_2012}%
  \BibitemOpen
  \bibfield  {author} {\bibinfo {author} {\bibfnamefont {J.-P.}\ \bibnamefont
  {Reid}}, \bibinfo {author} {\bibfnamefont {M.~A.}\ \bibnamefont {Tanatar}},
  \bibinfo {author} {\bibfnamefont {A.}~\bibnamefont {Juneau-Fecteau}},
  \bibinfo {author} {\bibfnamefont {R.~T.}\ \bibnamefont {Gordon}}, \bibinfo
  {author} {\bibfnamefont {S.~R.}\ \bibnamefont {de~Cotret}}, \bibinfo {author}
  {\bibfnamefont {N.}~\bibnamefont {Doiron-Leyraud}}, \bibinfo {author}
  {\bibfnamefont {T.}~\bibnamefont {Saito}}, \bibinfo {author} {\bibfnamefont
  {H.}~\bibnamefont {Fukazawa}}, \bibinfo {author} {\bibfnamefont
  {Y.}~\bibnamefont {Kohori}}, \bibinfo {author} {\bibfnamefont
  {K.}~\bibnamefont {Kihou}}, \bibinfo {author} {\bibfnamefont {C.~H.}\
  \bibnamefont {Lee}}, \bibinfo {author} {\bibfnamefont {A.}~\bibnamefont
  {Iyo}}, \bibinfo {author} {\bibfnamefont {H.}~\bibnamefont {Eisaki}},
  \bibinfo {author} {\bibfnamefont {R.}~\bibnamefont {Prozorov}}, \ and\
  \bibinfo {author} {\bibfnamefont {L.}~\bibnamefont {Taillefer}},\ }\href
  {\doibase 10.1103/PhysRevLett.109.087001} {\bibfield  {journal} {\bibinfo
  {journal} {Physical Review Letters}\ }\textbf {\bibinfo {volume} {109}},\
  \bibinfo {pages} {087001} (\bibinfo {year} {2012})}\BibitemShut {NoStop}%
\bibitem [{\citenamefont {Dong}\ \emph {et~al.}(2010)\citenamefont {Dong},
  \citenamefont {Zhou}, \citenamefont {Guan}, \citenamefont {Zhang},
  \citenamefont {Dai}, \citenamefont {Qiu}, \citenamefont {Wang}, \citenamefont
  {He}, \citenamefont {Chen},\ and\ \citenamefont {Li}}]{dong_quantum_2010}%
  \BibitemOpen
  \bibfield  {author} {\bibinfo {author} {\bibfnamefont {J.~K.}\ \bibnamefont
  {Dong}}, \bibinfo {author} {\bibfnamefont {S.~Y.}\ \bibnamefont {Zhou}},
  \bibinfo {author} {\bibfnamefont {T.~Y.}\ \bibnamefont {Guan}}, \bibinfo
  {author} {\bibfnamefont {H.}~\bibnamefont {Zhang}}, \bibinfo {author}
  {\bibfnamefont {Y.~F.}\ \bibnamefont {Dai}}, \bibinfo {author} {\bibfnamefont
  {X.}~\bibnamefont {Qiu}}, \bibinfo {author} {\bibfnamefont {X.~F.}\
  \bibnamefont {Wang}}, \bibinfo {author} {\bibfnamefont {Y.}~\bibnamefont
  {He}}, \bibinfo {author} {\bibfnamefont {X.~H.}\ \bibnamefont {Chen}}, \ and\
  \bibinfo {author} {\bibfnamefont {S.~Y.}\ \bibnamefont {Li}},\ }\href
  {\doibase 10.1103/PhysRevLett.104.087005} {\bibfield  {journal} {\bibinfo
  {journal} {Physical Review Letters}\ }\textbf {\bibinfo {volume} {104}},\
  \bibinfo {pages} {087005} (\bibinfo {year} {2010})}\BibitemShut {NoStop}%
\bibitem [{\citenamefont {Hashimoto}\ \emph {et~al.}(2010)\citenamefont
  {Hashimoto}, \citenamefont {Serafin}, \citenamefont {Tonegawa}, \citenamefont
  {Katsumata}, \citenamefont {Okazaki}, \citenamefont {Saito}, \citenamefont
  {Fukazawa}, \citenamefont {Kohori}, \citenamefont {Kihou}, \citenamefont
  {Lee}, \citenamefont {Iyo}, \citenamefont {Eisaki}, \citenamefont {Ikeda},
  \citenamefont {Matsuda}, \citenamefont {Carrington},\ and\ \citenamefont
  {Shibauchi}}]{hashimoto_evidence_2010}%
  \BibitemOpen
  \bibfield  {author} {\bibinfo {author} {\bibfnamefont {K.}~\bibnamefont
  {Hashimoto}}, \bibinfo {author} {\bibfnamefont {A.}~\bibnamefont {Serafin}},
  \bibinfo {author} {\bibfnamefont {S.}~\bibnamefont {Tonegawa}}, \bibinfo
  {author} {\bibfnamefont {R.}~\bibnamefont {Katsumata}}, \bibinfo {author}
  {\bibfnamefont {R.}~\bibnamefont {Okazaki}}, \bibinfo {author} {\bibfnamefont
  {T.}~\bibnamefont {Saito}}, \bibinfo {author} {\bibfnamefont
  {H.}~\bibnamefont {Fukazawa}}, \bibinfo {author} {\bibfnamefont
  {Y.}~\bibnamefont {Kohori}}, \bibinfo {author} {\bibfnamefont
  {K.}~\bibnamefont {Kihou}}, \bibinfo {author} {\bibfnamefont {C.~H.}\
  \bibnamefont {Lee}}, \bibinfo {author} {\bibfnamefont {A.}~\bibnamefont
  {Iyo}}, \bibinfo {author} {\bibfnamefont {H.}~\bibnamefont {Eisaki}},
  \bibinfo {author} {\bibfnamefont {H.}~\bibnamefont {Ikeda}}, \bibinfo
  {author} {\bibfnamefont {Y.}~\bibnamefont {Matsuda}}, \bibinfo {author}
  {\bibfnamefont {A.}~\bibnamefont {Carrington}}, \ and\ \bibinfo {author}
  {\bibfnamefont {T.}~\bibnamefont {Shibauchi}},\ }\href {\doibase
  10.1103/PhysRevB.82.014526} {\bibfield  {journal} {\bibinfo  {journal}
  {Physical Review B}\ }\textbf {\bibinfo {volume} {82}},\ \bibinfo {pages}
  {014526} (\bibinfo {year} {2010})}\BibitemShut {NoStop}%
\bibitem [{\citenamefont {Larson}\ and\ \citenamefont
  {Von~Dreele}(2000)}]{GSAS_2000}%
  \BibitemOpen
  \bibfield  {author} {\bibinfo {author} {\bibfnamefont {A.}~\bibnamefont
  {Larson}}\ and\ \bibinfo {author} {\bibfnamefont {R.}~\bibnamefont
  {Von~Dreele}},\ }\href
  {http://www.ccp14.ac.uk/ccp/ccp14/ftp-mirror/gsas/public/gsas/manual/GSASMan%
ual.pdf} {\bibfield  {journal} {\bibinfo  {journal} {Los Alamos National
  Laboratory Report LAUR 86-748}\ } (\bibinfo {year} {2000})}\BibitemShut
  {NoStop}%
\bibitem [{\citenamefont {Bud'ko}\ \emph {et~al.}(2013)\citenamefont {Bud'ko},
  \citenamefont {Sturza}, \citenamefont {Chung}, \citenamefont {Kanatzidis},\
  and\ \citenamefont {Canfield}}]{budko_heat_2013}%
  \BibitemOpen
  \bibfield  {author} {\bibinfo {author} {\bibfnamefont {S.~L.}\ \bibnamefont
  {Bud'ko}}, \bibinfo {author} {\bibfnamefont {M.}~\bibnamefont {Sturza}},
  \bibinfo {author} {\bibfnamefont {D.~Y.}\ \bibnamefont {Chung}}, \bibinfo
  {author} {\bibfnamefont {M.~G.}\ \bibnamefont {Kanatzidis}}, \ and\ \bibinfo
  {author} {\bibfnamefont {P.~C.}\ \bibnamefont {Canfield}},\ }\href {\doibase
  10.1103/PhysRevB.87.100509} {\bibfield  {journal} {\bibinfo  {journal}
  {Physical Review B}\ }\textbf {\bibinfo {volume} {87}},\ \bibinfo {pages}
  {100509} (\bibinfo {year} {2013})}\BibitemShut {NoStop}%
\bibitem [{\citenamefont {Murnaghan}(1937)}]{murnaghan_finite_1937}%
  \BibitemOpen
  \bibfield  {author} {\bibinfo {author} {\bibfnamefont {F.~D.}\ \bibnamefont
  {Murnaghan}},\ }\href {\doibase 10.2307/2371405} {\bibfield  {journal}
  {\bibinfo  {journal} {American Journal of Mathematics}\ }\textbf {\bibinfo
  {volume} {59}},\ \bibinfo {pages} {235} (\bibinfo {year} {1937})}\BibitemShut
  {NoStop}%
\bibitem [{\citenamefont {Thomale}\ \emph {et~al.}(2011)\citenamefont
  {Thomale}, \citenamefont {Platt}, \citenamefont {Hanke}, \citenamefont {Hu},\
  and\ \citenamefont {Bernevig}}]{thomale_exotic_2011}%
  \BibitemOpen
  \bibfield  {author} {\bibinfo {author} {\bibfnamefont {R.}~\bibnamefont
  {Thomale}}, \bibinfo {author} {\bibfnamefont {C.}~\bibnamefont {Platt}},
  \bibinfo {author} {\bibfnamefont {W.}~\bibnamefont {Hanke}}, \bibinfo
  {author} {\bibfnamefont {J.}~\bibnamefont {Hu}}, \ and\ \bibinfo {author}
  {\bibfnamefont {B.~A.}\ \bibnamefont {Bernevig}},\ }\href {\doibase
  10.1103/PhysRevLett.107.117001} {\bibfield  {journal} {\bibinfo  {journal}
  {Physical Review Letters}\ }\textbf {\bibinfo {volume} {107}},\ \bibinfo
  {pages} {117001} (\bibinfo {year} {2011})}\BibitemShut {NoStop}%
\bibitem [{\citenamefont {Hammersley}\ \emph {et~al.}(1996)\citenamefont
  {Hammersley}, \citenamefont {Svensson}, \citenamefont {Hanfland},
  \citenamefont {Fitch},\ and\ \citenamefont
  {Hausermann}}]{hammersley_two-dimensional_1996}%
  \BibitemOpen
  \bibfield  {author} {\bibinfo {author} {\bibfnamefont {A.~P.}\ \bibnamefont
  {Hammersley}}, \bibinfo {author} {\bibfnamefont {S.~O.}\ \bibnamefont
  {Svensson}}, \bibinfo {author} {\bibfnamefont {M.}~\bibnamefont {Hanfland}},
  \bibinfo {author} {\bibfnamefont {A.~N.}\ \bibnamefont {Fitch}}, \ and\
  \bibinfo {author} {\bibfnamefont {D.}~\bibnamefont {Hausermann}},\ }\href
  {\doibase 10.1080/08957959608201408} {\bibfield  {journal} {\bibinfo
  {journal} {High Pressure Research}\ }\textbf {\bibinfo {volume} {14}},\
  \bibinfo {pages} {235} (\bibinfo {year} {1996})}\BibitemShut {NoStop}%
\bibitem [{\citenamefont {Jeffries}\ \emph {et~al.}(2012)\citenamefont
  {Jeffries}, \citenamefont {Butch}, \citenamefont {Kirshenbaum}, \citenamefont
  {Saha}, \citenamefont {Samudrala}, \citenamefont {Weir}, \citenamefont
  {Vohra},\ and\ \citenamefont {Paglione}}]{jeffries_suppression_2012}%
  \BibitemOpen
  \bibfield  {author} {\bibinfo {author} {\bibfnamefont {J.~R.}\ \bibnamefont
  {Jeffries}}, \bibinfo {author} {\bibfnamefont {N.~P.}\ \bibnamefont {Butch}},
  \bibinfo {author} {\bibfnamefont {K.}~\bibnamefont {Kirshenbaum}}, \bibinfo
  {author} {\bibfnamefont {S.~R.}\ \bibnamefont {Saha}}, \bibinfo {author}
  {\bibfnamefont {G.}~\bibnamefont {Samudrala}}, \bibinfo {author}
  {\bibfnamefont {S.~T.}\ \bibnamefont {Weir}}, \bibinfo {author}
  {\bibfnamefont {Y.~K.}\ \bibnamefont {Vohra}}, \ and\ \bibinfo {author}
  {\bibfnamefont {J.}~\bibnamefont {Paglione}},\ }\href {\doibase
  10.1103/PhysRevB.85.184501} {\bibfield  {journal} {\bibinfo  {journal}
  {Physical Review B}\ }\textbf {\bibinfo {volume} {85}},\ \bibinfo {pages}
  {184501} (\bibinfo {year} {2012})}\BibitemShut {NoStop}%
\end{thebibliography}%

\end{document}